# Effect of doping of Co, Ni and Ga on magnetic and dielectric properties of layered perovskite multiferroic YBaCuFeO$_5$

Surender Lal*, C. S. Yadav and K. Mukherjee

School of Basic Sciences, Indian Institute of Technology Mandi, Mandi 175005, Himachal Pradesh, India

*Corresponding Author: Surender Lal

School of Basic Sciences,

Indian Institute of Technology Mandi,

Mandi 175005, Himachal Pradesh, India

Tel: +919418304396

E-mail: surenderlal30@gmail.com

# Abstract

YBaCuFeO$_5$ is one of the interesting multiferroic compounds, which exhibits magnetic ordering and dielectric anomaly above 200 K. Partial substitution of Fe with other magnetic and non-magnetic ion affects the magnetic and the structural properties of the system. We report detailed investigation of structural, magnetic and dielectric properties of YBaCuFe$_{0.85}$M$_{0.15}$O$_5$ (M=Co, Ni and Ga). We observed that the partial replacement of Ni and Co in place of Fe, results in magnetic dilution and broadening of the magnetic transition and shifting towards lower temperature. The replacement of Fe with non-magnetic Ga also results in shifting of the magnetic transition to the lower temperature side. The observed dielectric relaxation behavior in these compounds is due to the charge carrier hoping. This study highlights the impacts of magnetic and non-magnetic doping at the magnetic site on magnetic and dielectric properties in layered perovskite compound YBaCuFeO$_5$.

Key words: Magnetic properties, glassy dynamics, dielectric relaxation

## 1. Introduction

The coexistence of magnetic and ferroelectric ordering in the same crystallographic phase makes the multiferroic materials quite interesting [1-3]. These materials are classified in two categories: Type-I, and Type-II. In type-I multiferroics, the magnetic and dielectric transitions are far away and independent of each other. Therefore, the possibility of coupling between magnetic and electric order parameters is very weak. On the other hand, in type II materials, the dielectric ordering follows the magnetic ordering and there is a possibility of strong coupling between these two orders parameters [1]. Further, the electric polarization can be induced in these materials by magnetic field or vice versa and this phenomenon is known as magnetoelectric coupling [2,3]. In recent year the compounds belonging to LnBaM'M"$O_5$ family (where Ln is the rare-earth ions, and M', M" are the transition metal ions) have been widely studied. In such compounds, two layers of M'$O_5$ and M"$O_5$ are present and it is possible to combine different types of metal element in an ordered or disordered network. Also depending upon the size of rare-earth ions, these compounds have the various interesting physical properties [4–7]. YBaCuFe$O_5$ is quite interesting, as it exhibits magnetic and dielectric transitions above ~ 200 K [8]. This compound undergoes a paramagnetic to commensurate antiferromagnetic (CM-AFM) transition temperature below $T_{N1}$ ~ 440 K, followed by a commensurate to incommensurate antiferromagnetic (ICM-AFM) transition at $T_{N2}$~ 200 K [8]. Recent investigations revealed that the chemical pressure has influence on the magnetic and the dielectric properties due to the presence of cation (Cu/Fe) disorder in the system [9]. This disorder can be induced in the system by varying the preparation conditions [10]. Generally, from the viewpoint of technological application, the multiferroics with magnetoelectric coupling near room temperature are useful. For tuning transition, temperature of the compound, disorder-induced studies has surfaced as an enormously useful technique. The disorder in this compound can also be induced by the replacement of Y with some other rare earth ions and such studies have already been carried out. A detailed investigation on LaBaCuFe$O_5$ and LuBaCuFe$O_5$ have revealed contrasting physical properties [11]. Studies on LnBaCuFe$O_5$ (Ln = Nd, Yb, Gd and Ho) reveals magnetic transitions in the compounds is not visible, where moments due to the rare-earth ions dominate the effect arising out of Cu/Fe ions [12], however the nature of magnetic transition is clearly probed by the Neutron diffraction in the recent investigations in spite of paramagnetic signal [13]. Investigations on HoBaCuFe$O_5$, GdBaCuFe$O_5$ and YbBaCuFe$O_5$ revealed that the observed upturn in the heat capacity at low temperatures in these compounds is due to Schottky anomaly [12]. One of the effective ways to introduce disorder in YBaCuFeO$_5$ is through partial replacement of Fe by other transition metal ions. Doping the Fe-site will introduce random impurities, which will directly affect the interactions arising out of FeO$_5$/CuO$_5$ bipyramids and hence might cause an alteration in the physical properties of YBaCuFeO$_5$.

In this article we present a detailed study of structural, magnetic and dielectric properties of YBaCuFe$_{0.85}$M$_{0.15}$O$_5$ (where M=Co, Ni and Ga). We substituted Fe by both magnetic and non-magnetic ions because the latter (Ga) might leads to the structural deformation only whereas the former (Co, Ni) will result in the change in magnetic properties along with structural deformation. As compared with Fe, Ga has higher ionic radii whereas Co and Ni have the comparably smaller ionic radii, in its +3 oxidation state. Here we would like to mention that the maximum solubility of M in YBaCuFeO$_5$ is about 15% of Fe, as revealed by our investigations. The partial replacement of Fe by Co and Ni leads to broadening in magnetic transition temperature. Doping with non-magnetic ion Ga leads to the expansion in the unit cell and influences the magnetic ordering. The dielectric relaxation has also been observed in these compounds. However, the interactions between the electric dipoles are not strong enough for collective freezing of the electric dipoles. As compared to the parent compound, the change in magnetodielectric coupling is found to be insignificant in the doped compounds.

## 2. Experimental details

Polycrystalline samples of YBaCuFe$_{0.85}$Co$_{0.15}$O$_5$ (Co_0.15), YBaCuFe$_{0.85}$Ni$_{0.15}$O$_5$ (Ni_0.15) and YBaCuFe$_{0.85}$Ga$_{0.15}$O$_5$ (Ga_0.15) are synthesized by solid-state reaction method similar to that reported in literature [9,14]. Power x-ray diffraction is performed using Regaku smart lab diffractometer using monochromatized CuKα$_1$ radiation at room temperature. DC magnetization measurements are carried out in a magnetic property measurement system (Quantum Design USA). Hioki LCR meter is used for temperature dependence of dielectric constant measurements integrated with physical property measurement system (Quantum, Design USA) with a setup from Cryonano Labs. For electric measurements, silver paint contacts are made to the polycrystalline pellets with typical electrode area A = 15 mm$^2$ and thickness d= 0.506 mm.

## 3. Results and Discussion

### 3.1 Structural properties

Rietveld refined x-ray diffraction patterns of all the studied compounds are shown in figure 1 and calculated parameters are tabulated in table 1. These compounds crystallize in the tetragonal structure (space group: P4mm). All the peaks fit well with the theoretical curve and no impurity peak is observed which indicates that all the compounds are formed in crystallographic single phase. The structure of these compounds is similar to that reported in literature [9,15,16]. The partial replacement of Fe with Co, Ni and Ga results in small changes in the structural parameters of the unit cell in comparison to $YBaCuFeO_5$ due to variation in the ionic radii.

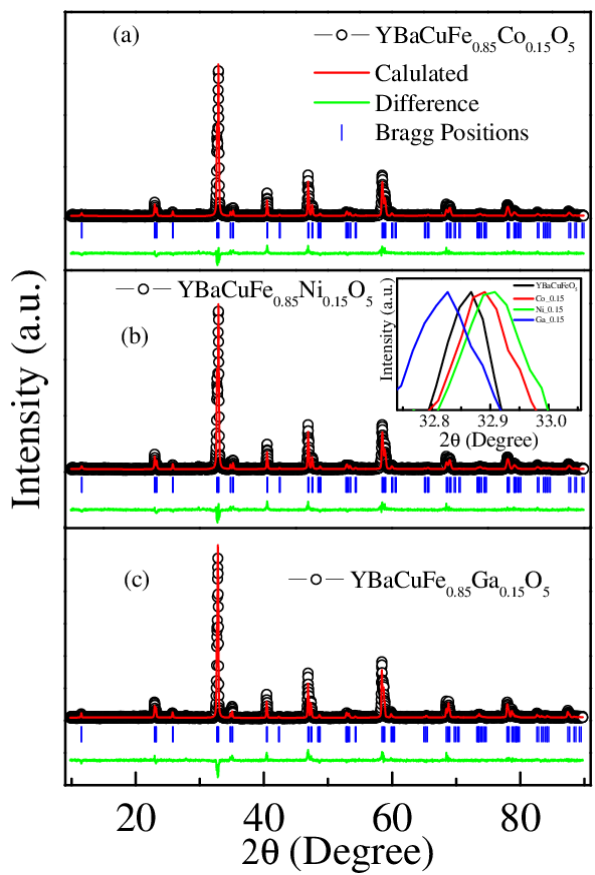

Figure 1: (a-c) Rietveld refined x-ray diffraction patterns of $YBaCuFe_{0.85}M_{0.15}O_5$ (M=Co, Ni, Ga) compounds. Inset of (b) shows the pattern in an expanded form for one peak for all the compounds, to bring out that the peaks shift with substitution. The data of the $YBaCuFeO_5$ is taken from Ref [9] for comparison.

Inset of figure 1 (b) show the shifting of x-ray diffraction peak with substitution. The x-ray diffraction curve of $YBaCuFeO_5$ for comparison is taken from Ref [9]. The peak position is shifted towards higher angle side with Co and Ni whereas Ga substitution shifts the peak position towards lower angle side; indicating that the former substitutions results in lattice contraction whereas the later one leads to lattice expansion. Co and Ni substitution leads to slight increase in *a*, but *c* lattice parameter decreases as compared to the YBaCuFeO$_5$. The partial replacement of Fe with Ga leads the increase in both *a,* and *c* parameters. The replaced ions sits at the Fe site at (1/2, 1/2, x) position. The distance between the pyramids in Co_0.15 and Ni_0.15 compounds is smaller as compared to the YBaCuFeO$_5$ whereas in Ga_0.15 it increases.

Table 1: Lattice parameter calculated from the Rietveld refinement of x-diffraction of YBaCuFe$_{0.85}$M$_{0.15}$O$_5$ (M=Co,Ni, Ga)

| Parameters | YBaCuFeO$_5$ [9] | YBaCuFe$_{0.85}$M$_{0.15}$O$_5$ | | |
|---|---|---|---|---|
| M | | Co_0.15 | Ni_0.15 | Ga_0.15 |
| a(Å) | 3.871 | 3.873(1) | 3.873(2) | 3.873(1) |
| c(Å) | 7.662 | 7.652(1) | 7.648(1) | 7.676(2) |
| V(Å$^3$) | 114.83 | 114.80 | 114.69 | 115.17 |
| R-factor | 4.92 | 9.20 | 11.8 | 17.7 |
| RF-factor | 4.18 | 8.23 | 15.1 | 13.1 |
| $\chi^2$ | 2.18 | 1.80 | 1.95 | 1.85 |
| Inter Pyramid distance | 2.833(1) | 2.829(2) | 2.827(3) | 2.838(2) |

## 3.2 Magnetic properties

The figure 2(a-c) shows the temperature dependence of the DC magnetic susceptibility of Co_0.15, Ni_0.15 and Ga_0.15 in the zero field cooled (ZFC) and field cooled (FC) conditions measured at 100 Oe. The figure 2(d-f) shows the temperature dependence of magnetization at 5 kOe of magnetic field. The commensurate to incommensurate antiferromagnetic transition is seen in the temperature range of 200-230 K for YBaCuFeO$_5$ compounds [8,17,18]. It is noted that the transition is shifted towards lower temperature ~ 200, 190 K and 110 K for Co_0.15, Ni_0.15 and Ga_0.15 respectively.

For Co_0.15 compound, a significant bifurcation between the ZFC and FC curves is noted below 100 K. Also as the temperature is decreased a weak kink is noted in the ZFC curve. Here, it is to be noted that no kink (at low temperatures) or significant bifurcation between ZFC and FC curves is observed in YBaCuFeO$_5$ [9]. For Ni_0.15 compound the bifurcation starts from 200 K and it becomes more prominent at lower temperature. A sharp peak at ~12.5 K is observed in the ZFC curve. In contrast, for Ga_0.15 compounds, no significant bifurcation between ZFC and FC curves is observed; however, a weak signature of a peak is present at ~11.5 K. Insets of figure 2(d-f) shows the *M* (*H*) at 2 and 300 K which show the linear behavior. It is to be noted that the unit cell volume of Ni_0.15 and Co_0.15 show very small change in comparison to YBaCuFeO$_5$. However, in contrast to Fe$^{3+}$ (which have five elections in 3$d$ shell), Co$^{3+}$ and Ni$^{3+}$ have six and seven electrons respectively. The Co$^{3+}$ have the three possible spin configurations, low spin, intermediate spin and high spin configuration [19]. The replacement of Fe$^{3+}$ by Co$^{3+}$ and Ni$^{3+}$ results in the broadening of magnetic transition and shifting of the transition towards the lower temperature. The observed behavior in the magnetization may be due to presence of different electronic states such as low spin and high spin configuration of Co$^{3+}$ and Ni$^{3+}$ ions. The partial replacement of Fe$^{3+}$ with Co$^{3+}$ or Ni$^{3+}$ seems to weakening the magnetic state and the magnetic interactions within the bipyramid. These states may give complex magnetic behavior and the magnetic transition is shifted to low temperature side. The substitution of Ga in place of Fe$^{3+}$ leads to expansion in the unit cell as compared to YBaCuFeO$_5$, which in turn increases the distance between the magnetic ions within the bipyramids. This changes the antiferromagnetic interactions within the unit cell and commensurate incommensurate antiferromagnetic transition shifts towards the lower temperature [11]. At low temperature, a kink/peak is observed in all the three compounds. This peak/kink observed at low temperature and low field is suppressed under higher applied field (shown in figure 2 (d-f)). Huge bifurcation between the ZFC and FC curve is noted below this kink/peak. This effect is more pronounced in Co, and Ni compounds. This bifurcation is more pronounced in the compounds with volume lower than the unit cell volume of YBaCuFeO$_5$ [11]. As stated before, Co and Ni substitution leads the decrease in lattice parameter $c$, thereby resulting in an increase of interaction between the magnetic ions. As the temperature is reduced, the crystal structures contracts and the magnetic interaction further changes, resulting in a possible change of magnetic anisotropy, which may contribute to the observation of these features. The suppression of these features with increasing the magnetic field also gives an indication that the origin of these characteristics is due to the weakening of the magnetic structure of the system. This feature in the ZFC and FC is less prominent in Ga_0.15 due to larger volume of the unit cell as compared to YBaCuFeO$_5$ due to the lattice expansion.

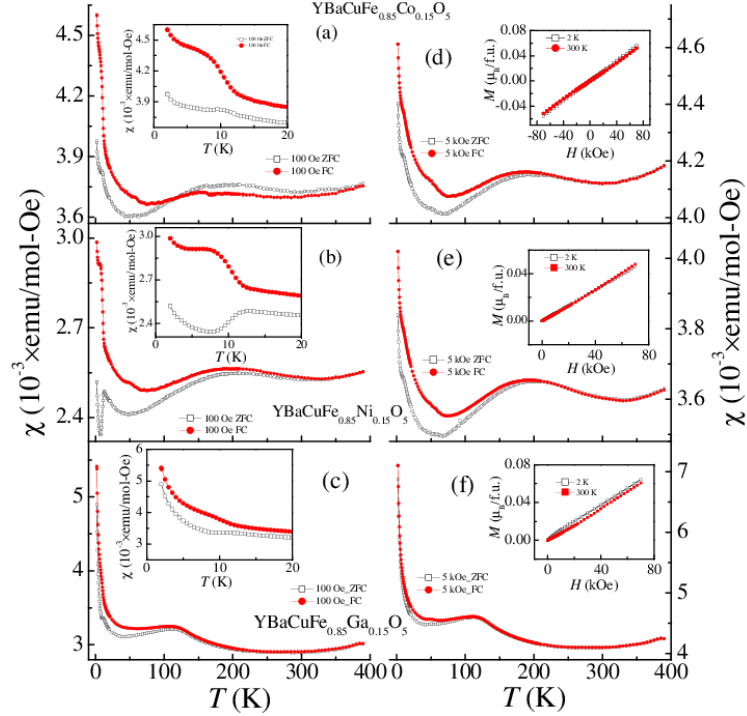

Figure 2: (a-c) Temperature dependence of *dc* susceptibility measured the ZFC and FC conditions for YBaCuFe$_{0.85}$M$_{0.15}$O$_5$ (M=Co, Ni and Ga) at 100 Oe of magnetic field and (d-f) at 5 kOe. Inset of (a-c) shows the *dc* susceptibility at low temperature. Inset (d-f) shows the isothermal response of magnetic field at 2 K and 300 K.

However, to resolve this new low temperature anomaly, temperature dependent neutron diffraction is warranted.

### 3.3 Dielectric analysis

The temperature variation of real ($\varepsilon'$) and imaginary part ($\varepsilon''$) of dielectric constant ($\varepsilon'$) measured at selected frequencies for the series YBaCuFe$_{0.85}$M$_{0.15}$O$_5$ (M= Co, Ni, Ga) in the temperature range of 10 to 300 K (shown in figure 3). The temperature variation of $\varepsilon'$ for Co_0.15, Ni_0.15 and Ga_0.15 shows the sharp increase near 56, 105 and 62 K respectively at 10 kHz followed by a plateau in the high temperature region. It is noted that value of $\varepsilon''$ also increases in this temperature range. Interestingly, for this series, a frequency dependent behavior of $\varepsilon'$ and $\varepsilon''$ is observed. To analyze this feature, temperature derivative of $\varepsilon'$ is plotted as function of temperature. The curve shows a peak (not shown), the temperature of which increases with the increase in frequency. In compounds where electric dipoles exhibits glassy dynamics such shift in the peak temperature are observed.

This variation of peak temperature can be analyzed by Arrhenius, Vogel Fulcher and/or power law [20]. We tried to fit the data with the above-mentioned laws. In our case the best fit is obtained with the Arrhenius law [21] of the form

$$\tau = \tau_0 \exp\left[\frac{E_a}{K_B T}\right] \quad \ldots (1)$$

Where $\tau_0$ is the pre-exponential factor and $E_a$ is the activation energy and $k_B$ is the Boltzmann constant. Upper Insets of figure 3 (a-c) shows the temperature variation of $\ln\tau$. The obtained values of the $E_a$ are 0.044 eV, 0.147 eV, 0.0729 eV $8.70\times10^{-9}$ s and $\tau_0$ are $1.09\times10^{-11}$ s, $4.707\times10^{-10}$ s for Co_0.15, Ni_0.15 and Ga_0.15 respectively. The values of $E_a$ are smaller as compared to Bi-

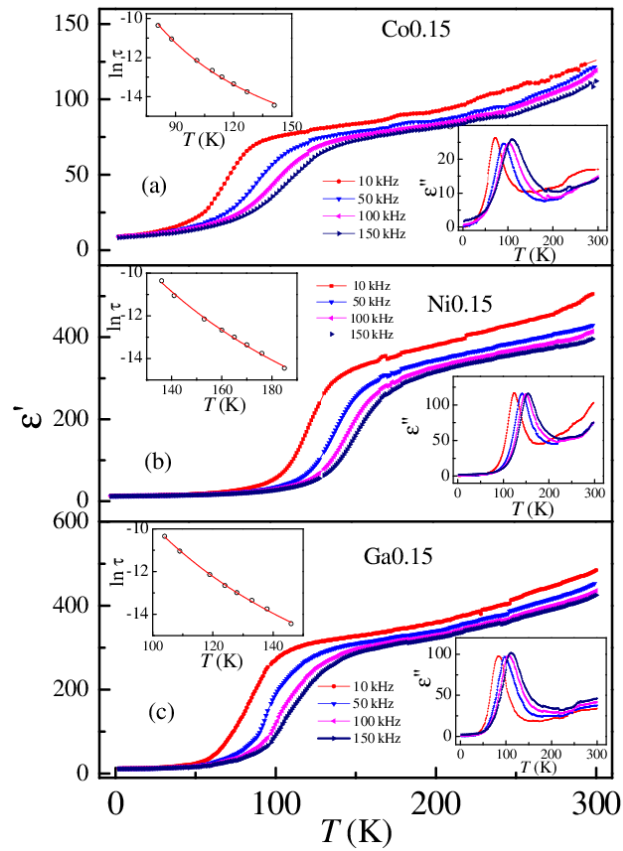

Figure 3: Temperature dependence of dielectric susceptibility in the temperature range of 2 -300 K of YBaCuFe$_{0.85}$M$_{0.15}$O$_5$ (M= Co, Ni and Ga). The lower inset shows imaginary part of dielectric constant and the upper inset shows the Arrhenius fit of the peak temperature obtained from the derivative of the real part of the dielectric constant.

Doped $SrTiO_3$ (0.74 to 0.86 eV) [22], $Bi_4Ti_3O_{12}$ (0.87 eV) [23] and $Bi_5TiNbWO_{15}$ (0.76 eV) [24] where the relaxation mechanism is ascribed to the thermal motion of oxygen vacancies. However, the values of $E_a$ is comparable to that observed for $La_2CoIrO_6$ (0.056 eV) [25], $Ca_3Co_{1.4}Rh_{0.6}O_6$ (0.071 eV) [26], $BiMn_2O_5$ (0.065 eV) [27] and $PrFe_{0.5}Mn_{0.5}O_{2.9}$ (0.19 eV) [28] where the relaxation behavior is due to the charge carrier hoping. The values of the activation energy of the studied compounds are comparable to the values observed for the charge carrier hoping. The relaxation in electric dipoles is due to the charge carrier hoping among the transition metal ion. However, the interaction between the electric dipoles is not strong enough for collective freezing of the electric dipoles. The partial replacement of Ba with Sr leads the dipolar glass behavior at low temperature [9]. However, this low temperature behavior is absent in Co, Ni and Ga doped compounds. This may be due to the effect of the transition metal ion, which may leads to the change in the interaction and affecting the collective freezing of electric dipoles.

### 3.4 Magneto-dielectric properties

$YBaCuFeO_5$ shows the magneto-dielectric coupling at different temperatures [8,9]. To see the effect of MDE coupling due to these substitutions, the magnetic field response of the dielectric constant is measured at 200 K at a fixed frequency of 50 kHz. The MDE effect is defined as $\Delta\varepsilon'(\%) = 100 \times [\varepsilon'(H)-\varepsilon'(0)]/\varepsilon'(0)$, where $\varepsilon'(H)$ and $\varepsilon'(0)$ are the dielectric constant in the presence and the absence of magnetic field respectively [9,29,30]. Figure 4 shows the magnetic field response of $\Delta\varepsilon'(\%)$. The data for $YBaCuFeO_5$ is taken from Ref [9] for comparison. It is observed that MDE coupling persists in all the doped compounds. However, the change in magnitude of $\Delta\varepsilon'(\%)$ is insignificant when compared with $YBaCuFeO_5$.

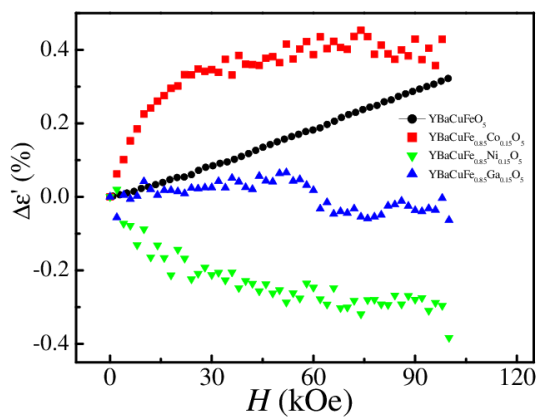

Figure 4: The magnetic field response of relative dielectric permittivity measured at different compounds measured at 200 K.

## 4. Summary

In summary, we have investigated the structural, magnetic dielectric and magneto-dielectric properties of YBaCuFe$_{0.85}$M$_{0.15}$O$_5$ (M=Co, Ni and Ga). Partial replacement of Fe with Co and Ni leads to broadening in the magnetic transition may be due to the change in magnetic structure of system. Doping with the non-magnetic ion shifts the magnetic transition towards the low temperature side as compared to YBaCuFeO$_5$. The dielectric relaxation behavior is observed. The observed may be due to the charge carrier hoping. However, glass-like behavior of electric dipoles in not observed, as the interaction between the electric dipoles are not strong enough. In addition, it is noted that the change in magnetodielectric coupling is insignificant in these doped compounds when compared with YBaCuFeO$_5$.

**Acknowledgement**

The authors acknowledge IIT Mandi for providing the experimental facilities. SL acknowledges the UGC India for SRF Fellowship. KM acknowledges the financial support from the CSIR project No. 03(1381)/16/EMR-II.